\documentclass[
aps,%
12pt,%
final,%
notitlepage,%
oneside,%
onecolumn,%
nobibnotes,%
nofootinbib,
superscriptaddress,%
showpacs,%
centertags]%
{revtex4}
\usepackage{amsmath}
\usepackage{epsfig}
\usepackage{citesort}
\usepackage{float}

\begin{document}
%\selectlanguage{english}
\title
{Kepler motion on single-sheet hyperboloid}

%----------------------------  AUTHORS  ---------------------------%
\author{Yu. A. Kurochkin}
\email {yukuroch@bas-net.by}
\author{V.S. Otchik}
\email{votchyk@tut.by}
\affiliation{B.I. Stepanov Institute of Physics, Minsk, Belarus}
\author{L. G. Mardoyan}
\email {mardoyan@ysu.am}
\affiliation{Joint Institute for Nuclear Research, Dubna, Russia}
\affiliation{Yerevan State University, Yerevan, Armenia}
\author{D.R. Petrosyan}
\email{petrosyan@theor.jinr.ru}
\affiliation{Joint Institute for Nuclear Research, Dubna, Russia}
\author{G.S. Pogosyan}
\email{pogosyan@theor.jinr.ru}
\affiliation{Joint Institute for Nuclear Research, Dubna, Russia}
\affiliation{Departamento de Matematicas, CUCEI,
Universidad de Guadalajara,  Guadalajara, Mexico}
\affiliation{Yerevan State University, Yerevan, Armenia}

\begin{abstract}

The classical Kepler-Coulomb problem on the single-sheeted hyperboloid  $H^{3}_1$
is solved in the framework of the Hamilton--Jacobi equation. We have proven that all the bounded orbits are closed
and periodic.  The paths are ellipses or circles for finite motion.
\end{abstract}


\maketitle
%#########################################################
\section{Introduction} \label{sec:INTRO}
%#########################################################

Classical and quantum mechanical systems in the spaces of constant curvature (positive and negative) have always drawn a lot of
attention due to connection with the relativistic physics and gravity. The 2D and 3D (pseudo)-Riemannian spaces of constant curvature are the models of the relativistic space-times (including de Sitter and anti de Sitter spaces) and find application in many
branches of physics, from the quantum gravity and cosmology \cite{HOOFT, Gibbons} to the quantum Hall effect \cite{HALL} or coherent
state quantization \cite{GAZEAU}.

The problem of the motion of a classical particle in the field of gravity and of a charged particle in the Coulomb field in the
spaces of constant curvature as in Euclidean space has a rich history. The introduction of hyperbolic geometry into the law of
gravity can be found already in the work of Lobachevsky, who was the first to define the Kepler potential and found the trajectory
of the classical motion \cite{chern} (see also the more recent articles devoted to the Kepler's problem on the three-dimensional sphere
and pseudosphere, i.e. Lobachevsky space \cite{KOZLOV1,KOZLOV2,DOMBROWSKI,SLAW}).
The investigation of Kepler--Coulomb problem in quantum mechanics was motivated by wish to compare the properties of the Coulomb potential
in the ``open hyperbolic'' or ``closed'' universe to that of an ``open but flat'' universe. Schr\"odinger \cite{SCHRO} was the first
who discussed this problem and discovered that for ``hydrogen atom'' on the three-dimensional sphere only discrete spectrum
exists. Later, Infeld and Schild \cite{INFELD} found that in an open hyperbolic universe there is only a finite (but very large)
number of bound states. Let us also mention some articles devoted to the investigation of various aspects of Kepler-Coulomb
problem in space of constant curvature, for instance \cite{HIG,BOGUSH-1,BOGUSH-2,BAR-2,GROSCH1,GROPOa,GROP5,GPSs,GPSs1}.

The motion in Coulomb field on the single-sheet hyperboloid, as shown by Grosche \cite{GROaa}
(see also \cite{KUROCH1}), has some peculiarities. The potential is not singular, and the discrete energy spectrum
is infinitely degenerate. Quite recently in the articles \cite{PETPOG1,PETPOG2} the classical and quantum
Kepler--Coulomb problem on the anti de Sitter $ AdS_3$ configuration space have been discussed. In particular it has been
shown in \cite{PETPOG2} that for the classical motion, like the cases of flat Euclidean space, sphere, and pseudosphere, all bounded trajectories are closed and periodic.

The present paper aims at investigating the classical Kepler problem in hyperbolic space $ H_1^3 = SO(3,1) / SO(2,1)$
(single-sheeted hyperboloid) which, to our knowledge, has not been elucidated in literature so far.

The classical Kepler problem on the single-sheeted hyperboloid $H_1^3: \, -x_0^2+x_1^2+x_2^2+x_3^2=R^2$,
where $x_i$ ($i=0,1,2,3$) are the Cartesian coordinates in the four-dimensional Minkowski space $M_{3,1}$,
is defined by the potential \cite{GROaa}
\begin{equation}
\label{EQUATION-1}
V = - \frac{\alpha}{R} \, \frac{x_0}{\sqrt{x_1^2+x_2^2+x_3^2}} = - \frac{\alpha}{R} \tanh\tau
\end{equation}
where $\tau$ is the ``quasi-radial angle''  corresponding to the geodesic pseudo-spherical coordinate on
$H_1^3$, namely
\begin{equation}
\label{PSEUDO:2}
x_0= R\sinh\tau, \,
x_1= R\cosh\tau\sin\theta\cos\phi, \,
x_2= R\cosh\tau \sin\theta\sin\phi, \,
x_3= R\cosh\tau \cos\theta,
\end{equation}
$$
\tau \in (-\infty, \infty), \qquad  \theta \in [0, \pi], \qquad  \phi \in [0, 2\pi).
$$
The standard metric of Minkowski space $M_{3,1}$: $ds^2= dx_0^2 - dx_1^2 - dx_2^2 - dx_3^2$ induces on
$H_1^3$ the indefinite metric
\begin{equation}
\label{METRIX-1}
\frac{ds^2}{R^2} =  d\tau^2 - \cosh^2\tau \left(d\theta^2 + \sin^2\theta d\phi^2\right).
\end{equation}
Then the kinetic energy is given by
\begin{equation}
\label{LAGRANGE-02}
T  = \frac{R^2}{2}\left[\dot{\tau}^2 - \cosh^2 \tau \left(\dot{\theta}^2 +
\sin^2\theta \dot{\phi}^2\right)\right],
\end{equation}
and the canonical momenta can be obtained in a usual way
\begin{equation}
\label{LAGRANGE-03}
p_\tau = \frac{\partial{T}}{\partial\dot{\tau}} = R^2 \dot{\tau},
\quad
p_\theta = \frac{\partial{ T}}{\partial\dot{\theta}} = - R^2\cosh^2 \tau \dot{\theta},
\quad
p_\phi = \frac{\partial{T}}{\partial\dot{\phi}} = - R^2 \cosh^2 \tau \sin^2 \theta \dot{\phi}.
\end{equation}
Thus the Hamiltonian describing the Kepler problem in the pseudo-spherical phase space
$(\tau, \theta, \phi;$ $ p_\tau, p_\theta, p_\phi)$ with respect to the canonical Poisson brackets
\begin{equation}
\label{CANONIC-01}
\{f, g \}  = \sum_{i=1}^{3}\left(\frac{\partial f}{\partial q_i}\frac{\partial g}{\partial p_i}
- \frac{\partial g}{\partial q_i}\frac{\partial f}{\partial p_i} \right)
\end{equation}
has the form
\begin{equation}
\label{eq:1}
 H  = \frac{1}{2 R^2}\left\{p_\tau^2 - \frac{1}{\cosh^2 \tau}
\left(p_\theta^2 +  \frac{p_\phi^2}{\sin^2\theta}\right)\right\}  - \frac{\alpha}{R} \tanh\tau.
\end{equation}
The Hamilton--Jacobi equation associated with the Hamiltonian (\ref{eq:1}) is obtained after the substitution
$p_{\mu_i} \to \partial S/ \partial\mu_i$, where $\mu_i = (\tau, \theta, \phi)$:
\begin{equation}
\label{H-J1}
\frac{1}{2R^2}\left\{\left(\frac{\partial S}{\partial \tau}\right)^2 - \frac{1}{\cosh^2\tau}
\left(\frac{\partial S}{\partial \theta}\right)^2 -
\frac{1}{\cosh^2\tau\sin^2\theta}\left(\frac{\partial S}{\partial \phi}\right)^2\right\}
- \frac{\alpha}{R} \tanh\tau  = E.
\end{equation}
This equation is completely separable, and the coordinate $\phi$ is cyclic. We seek solutions for the classical action
$S(\tau,\, \theta,\, \phi,\, t)$ in the form
\begin{equation}
\label{H-J2}
S(\tau, \theta, \phi, t) = -Et +  p_\phi \phi + S_1(\tau) + S_2(\theta),
\end{equation}
and obtain from (\ref{H-J1}) two ordinary differential equations
\begin{eqnarray}
\label{H-J3}
\left(\frac{d S_2}{d \theta}\right)^2 + \frac{p^2_\phi}{\sin^2\theta} &=& L^2,
\\[3mm]
\label{H-J3-1}
\left(\frac{d S_1}{d \tau}\right)^2 +  2R^2 {V}(\tau)   -
\frac{L^2}{\cosh^2\tau}& =& 2 R^2 E,
\end{eqnarray}
where $L^2$ is the separation constant which coincides with the square of the angular
momentum.
From equation (\ref{H-J3}) it follows that for a fixed $L^2 \geq 0$ the separation constant $p^2_\phi$
is in the range $0 \leq p^2_\phi \leq L^2$.
The equation (\ref{H-J3-1}) describes the motion in the field of the effective potential (see also Fig.1)
\begin{equation}
\label{H-EFF-0}
V_{eff}(\tau) = -  \frac{\alpha}{R} \tanh \tau -  \frac{L^2}{2R^2\cosh^2\tau}.
\end{equation}
In the case of $L^2 > \alpha R$ this potential has a minimum at
$\tau_0 = \tanh^{-1}[{\alpha R}/L^2]$ and at this point
\begin{equation}
\label{EQUATION-1.4}
V_{eff}(\tau_0) = - \frac{1}{2 R^2} \left(L^2 + \frac{\alpha^2 R^2}{L^2}\right).
\end{equation}

Using now equations (\ref{H-J3}) and (\ref{H-J3-1}) we get
\begin{eqnarray}
\label{H-J4-1}
S_1(\tau)&=& \int\sqrt{2R^2E - 2R^2 {V}(\tau) + \frac{L^2}{\cosh^2\tau}}\,d\tau,
\\[3mm]
\label{H-J4-2}
S_2(\tau) &=& \int\sqrt{L^2 - \frac{p^2_\phi}{\sin^2\theta}}\,d\theta.
\end{eqnarray}
Since only the trajectories are in the scope of our interest, we will follow the usual
procedures \cite{LL} and consider the equations
\begin{eqnarray}
\label{H-J5}
\frac{\partial S}{\partial E} = \frac{\partial S_1}{\partial E} - t = - t_0,
\quad
\frac{\partial S}{\partial L^2}= \frac{\partial S_1}{\partial L^2} +
\frac{\partial S_2}{\partial L^2} = \beta,
\quad
\frac{\partial S}{\partial
p_\phi} = \phi +  \frac{\partial S_2}{\partial
p_\phi}= \phi_0,
\end{eqnarray}
where $t_0$, $\phi_0$, and $\beta$ are constants.

\begin{figure}[H]
\centering
\includegraphics[scale=0.5]{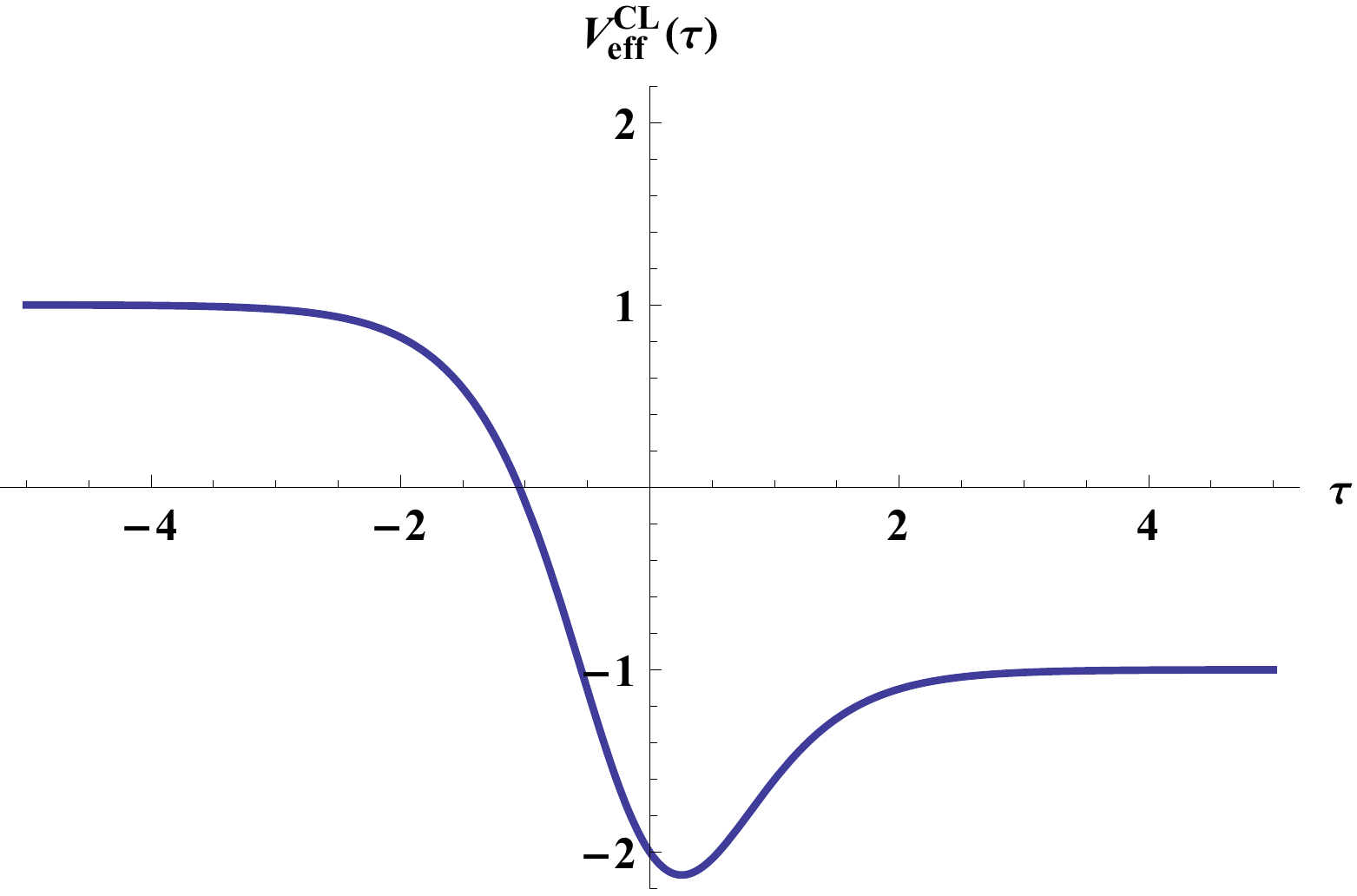}
\includegraphics[scale=0.5]{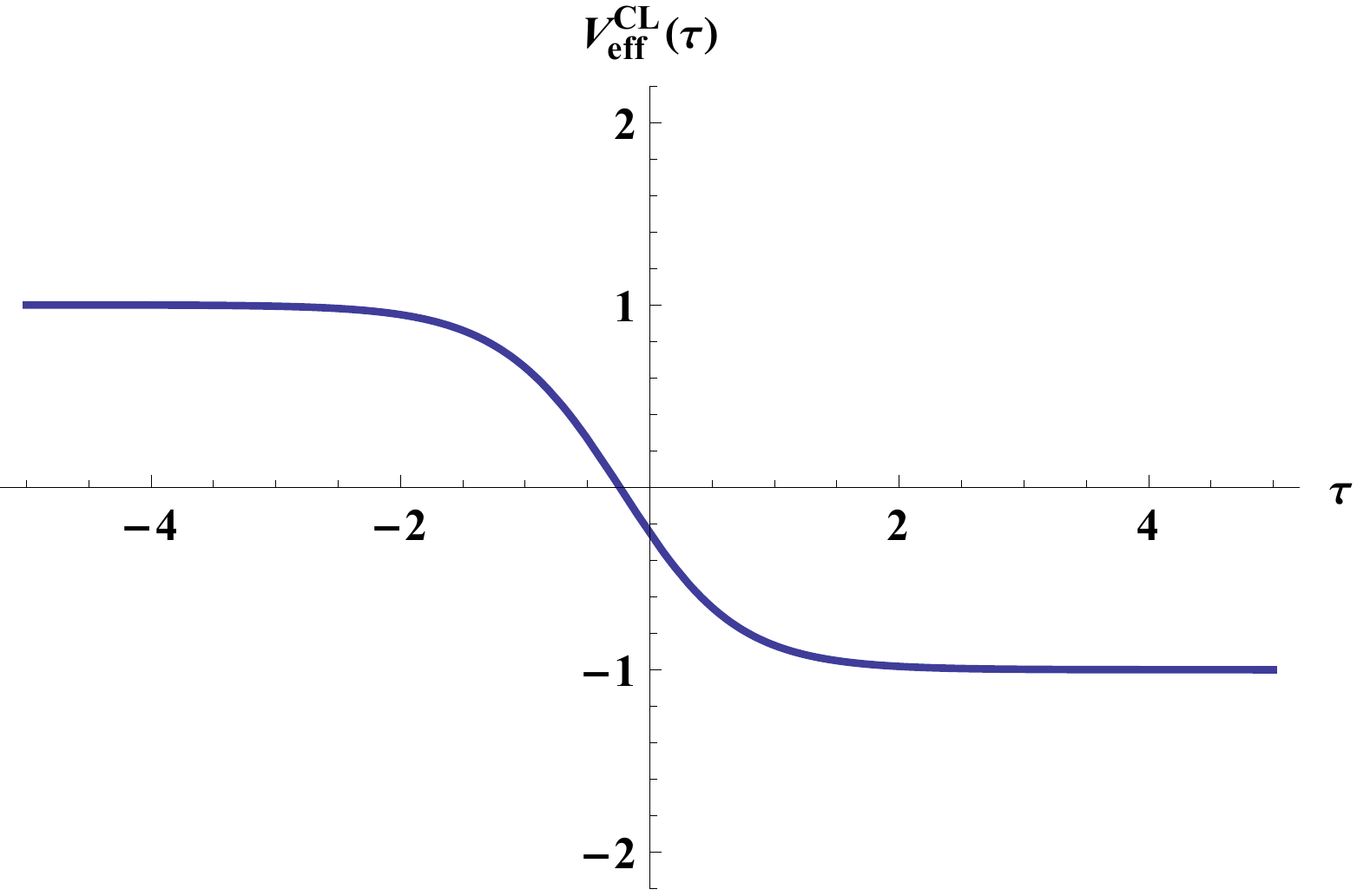}
\caption{$V^{Cl}_{eff}(\tau) = -  \frac{\alpha}{R} \tanh \tau -  \frac{L^2}{2R^2\cosh^2\tau}$
for $L^2=4$ and $L^2=1/2$ ($\alpha=R=1$)}
\label{fig:graphO21}
\end{figure}

From (\ref{H-J4-1})
and the first of equations (\ref{H-J5}) we obtain
\begin{equation}
\label{H-J6}
t-t_0 =  \frac{R^2}{L} \int\frac{d X}{(1-X^2) \sqrt{- X^2  +
{2\alpha R}/{L^2} X +  {2R^2E}/{L^2} + 1}}\,,
\end{equation}
where $X = \tanh \tau \in [-1, 1]$, and the roots of the subradical polynom in denominator are
\begin{equation}
\label{H-J8}
X_{1,2} = \frac{\alpha R}{L^2} \pm \sqrt{\left(\frac{\alpha R}{L^2}\right)^2
+ \frac{2R^2E}{L^2} +1}\,.
\end{equation}
The finite motion, when the range of $\tau$ is given by
\begin{equation}
\label{H-J8-1}
\frac{\alpha R}{L^2} - \sqrt{\left(\frac{\alpha R}{L^2}\right)^2
+ \frac{2R^2E}{L^2} +1}
< \tanh\tau <
\frac{\alpha R}{L^2} + \sqrt{\left(\frac{\alpha R}{L^2}\right)^2
+ \frac{2R^2E}{L^2} +1}\,,
\end{equation}
exists for the negative values of $E$ in the range
\begin{equation}
\label{H-J9}
- \frac{1}{2 R^2} \left(L^2 + \frac{\alpha^2 R^2}{L^2}\right) \leq E < - \frac{\alpha}{R}.
\end{equation}
In the case of $E=E_{\min} = - \left (L^2 + \alpha^2 R^2/L^2\right)/(2 R^2)$, the integral in (\ref{H-J6}) is not well defined and we turn directly to the equation (\ref{H-J3-1}). Then from equation
\begin{eqnarray}
\label{H-J3-3}
\left(\frac{d S_1}{d \tau}\right)^2 =
-  L^2 \left(\tanh \tau - \frac{\alpha R}{L^2}\right)^2
\end{eqnarray}
it follows that ${d S_1}/{d \tau}=0$ and $\tanh \tau = \alpha R/L^2$. Therefore
\begin{equation}
\label{H-J10}
\tau = \tanh^{-1}  \left(\frac{R|E|}{\alpha} + \sqrt{\frac{R^2 E^2}{\alpha^2}-1}\right),
\end{equation}
i.e. the paths are circles. In the case of $ - \alpha/R\leq E \leq \alpha/R$, the range in which $\tau$ may vary
is determined by $-1 +  \alpha R/L^2 \leq \tanh\tau \leq 1$, that is the motion is
limited from the left side but not limited from the right side, and particle can move
to the infinity. In the case of $E \geq \alpha/R$ we have $X_1=1+2\alpha R/L^2$, $X_2= -1$
and $-\infty < \tau < \infty$.

From equations (\ref{H-J4-1}) and (\ref{H-J4-2}) we obtain
\begin{multline}
\label{H-J14}
\frac{\partial S_1}{\partial L^2}
=  \frac{1}{2}\int\frac{\cosh^{-2}\tau d\tau}{\sqrt{2R^2E + 2\alpha R \tanh \tau - L^2/\cosh^2\tau}}=\\[2mm]
=  \frac{1}{2 L}
\, \arcsin \frac{\tanh\tau-\alpha R/L^2}{\sqrt{(\alpha R/L^2)^2 + 2R^2E/L^2 + 1}},
\end{multline}
\begin{equation}
\label{H-J15}
\frac{\partial S_2}{\partial L^2}
=  \frac12\int\frac{d\theta}{\sqrt{L^2 - p_\phi^2/\sin^2\theta}}
= \frac{1}{2L}
\, \arcsin \frac{\cos\theta}{\sqrt{1-p_{\phi}^2/L^2}}.
\end{equation}
In addition to equation (\ref{H-J9}) we require
$ -\sqrt{1-p_{\phi}^2/L^2} \leq \cos \theta \leq \sqrt{1- p_{\phi}^2/L^2}$ and
\begin{equation}
\label{H-J19}
|\tanh \tau - \alpha R/L^2| \leq \sqrt{(\alpha R/L^2)^2 + 2R^2E/L^2+1}.
\end{equation}
The last condition coincides with (\ref{H-J8-1}).
Finally for ${\partial S}/{\partial L^2}$ we have
\begin{equation}
\label{H-J20}
\frac{\partial S}{\partial L^2} = \frac{1}{2 L}
\left[\arcsin{\frac{\tanh \tau - \alpha R/L^2}{\sqrt{(\alpha R/L^2)^2 + 2R^2E/L^2+1}}}
+ \arcsin{\frac{\cos\theta}{\sqrt{1 - p_\phi^2/L^2}}}\right]
= \beta.
\end{equation}
At the next step, from equations (\ref{H-J4-2}) and (\ref{H-J5})  we obtain
\begin{eqnarray}
\label{H-J22}
\frac{\partial S}{\partial p_\phi}  = \phi + \int\frac{p_\phi
d\theta}{\sin^2\theta \sqrt{L^2 - p_\phi^2/\sin^2\theta}} =
\phi + \arcsin\frac{\cot\theta}{\sqrt{L^2/p_\phi^2-1}}
= \phi_0,
\end{eqnarray}
for $|\cot\theta| \leq  \sqrt{L^2/p_\phi^2-1}$ and hence
\begin{eqnarray}
\label{ANGL-J22}
\cot\theta = \sqrt{L^2/p_\phi^2-1}\,\sin(\phi_0-\phi).
\end{eqnarray}
It is easy to see that for the fixed values of $L^2$ and $p_\phi^2$ the motion of particle on the single-sheeted
hyperboloid is restricted by the additional condition
\begin{eqnarray}
\frac{x_3}{x_1 \sin\phi_0 - x_2 \cos\phi_0} = \sqrt{L^2/p_\phi^2-1}.
\end{eqnarray}
Therefore, without the loss of generality we can choose $L^2 = p_\phi^2$ or, which is the same, $\theta=\pi/2$, and $x_3 = 0$.
Thereby the path of the motion lies on the two-dimensional single-sheeted hyperboloid: $x_0^2-x_1^2-x_2^2 = - R^2$.
Eliminating the dependence on $\theta$ in equation (\ref{H-J20}) with the help of (\ref{ANGL-J22}),
and putting $L^2= p_\phi^2$, we can write the equation of the path in the form
\begin{equation}
\label{H-J29}
\frac{1}{1-\tanh \tau} =  \frac{p}{1 + \varepsilon(R) \, \cos \phi},
\end{equation}
where we use the notations
\begin{equation}
\label{H-J30}
p = \left(1-\frac{\alpha R}{L^2}\right)^{-1} > 0,
\qquad
\varepsilon (R) = \sqrt{1 +  \frac{2\alpha R}{L^2}\,
\frac{(1+ R E/\alpha)}{(1-\alpha R/L^2)^2}},
\end{equation}
and choose $\phi_0 = 2\beta L + \frac{\pi}{2}$ so that the point $\phi = 0$
will be the nearest to the center. When $0 \leq \varepsilon (R)< 1$, the orbits are ellipses (circles for
$\varepsilon (R) =0$), when $\varepsilon (R) = 1$, the path is a parabola and
when $\varepsilon (R) > 1$ the path is a hyperbola.

Now we summarize some properties of particles orbits.

{\bf A.} The elliptic orbit is only possible if
 $- \frac{1}{2 R^2} \left (L^2 + \frac{\alpha^2 R^2}{L^2}\right) < E < - \frac{\alpha}{R}$, and then
\begin{equation}
\label{H-J1-30}
0 \leq \varepsilon (R) =  \sqrt{1 +  \frac{2\alpha R}{L^2}\,
\frac{(1- R |E|/\alpha)}{(1-\alpha R/L^2)^2}} < 1.
\end{equation}
%---------------------------------------------------------------------------------------
Let $\tau_{\min}$ and $\tau_{\max}$ denote, respectively, the minimal and maximal angular
distances from the center of field. It is obvious that they correspond to the angles
$\phi=0$ and $\phi=\pi$. Accordingly to equations (\ref{H-J29}) and (\ref{H-J30}) we have
%---------------------------------------------------------------------------------------
\begin{eqnarray}
\label{H-J33}
\tau_{\min}
&=& \tanh^{-1} \left\{\frac{\alpha R}{L^2} -
\sqrt{\left(\frac{\alpha R}{L^2}\right)^2 - \frac{2R^2|E|}{L^2} + 1}\right\},
\\
\tau_{\max} &=& \tanh^{-1} \left\{\frac{\alpha R}{L^2} +
\sqrt{\left(\frac{\alpha R}{L^2}\right)^2 - \frac{2R^2|E|}{L^2} + 1}\right\}.
\end{eqnarray}

\vspace{0.2cm}

\begin{figure}[H]
\centering
\includegraphics[width=7cm,height=7cm]{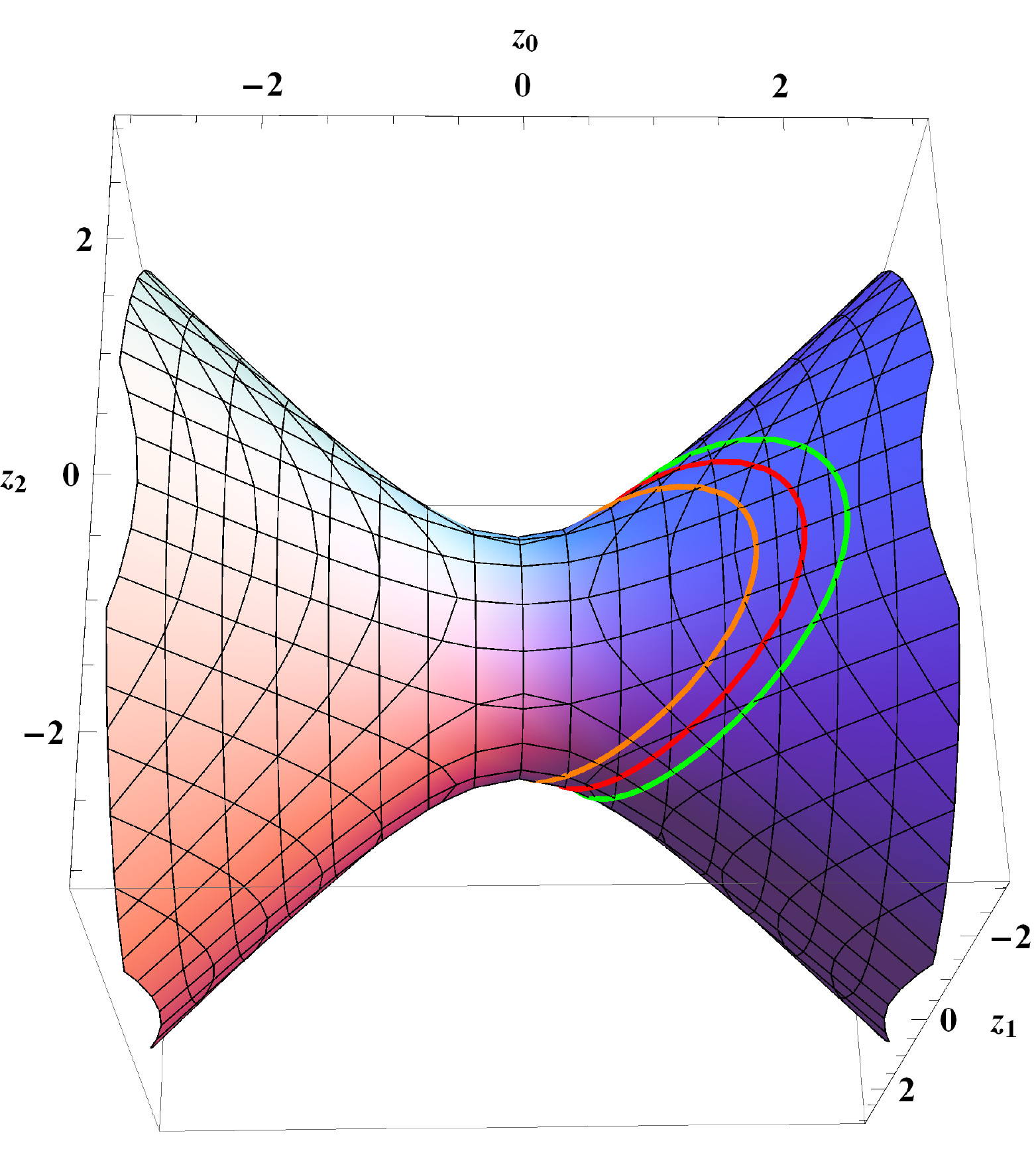}
\includegraphics[width=7cm,height=7cm]{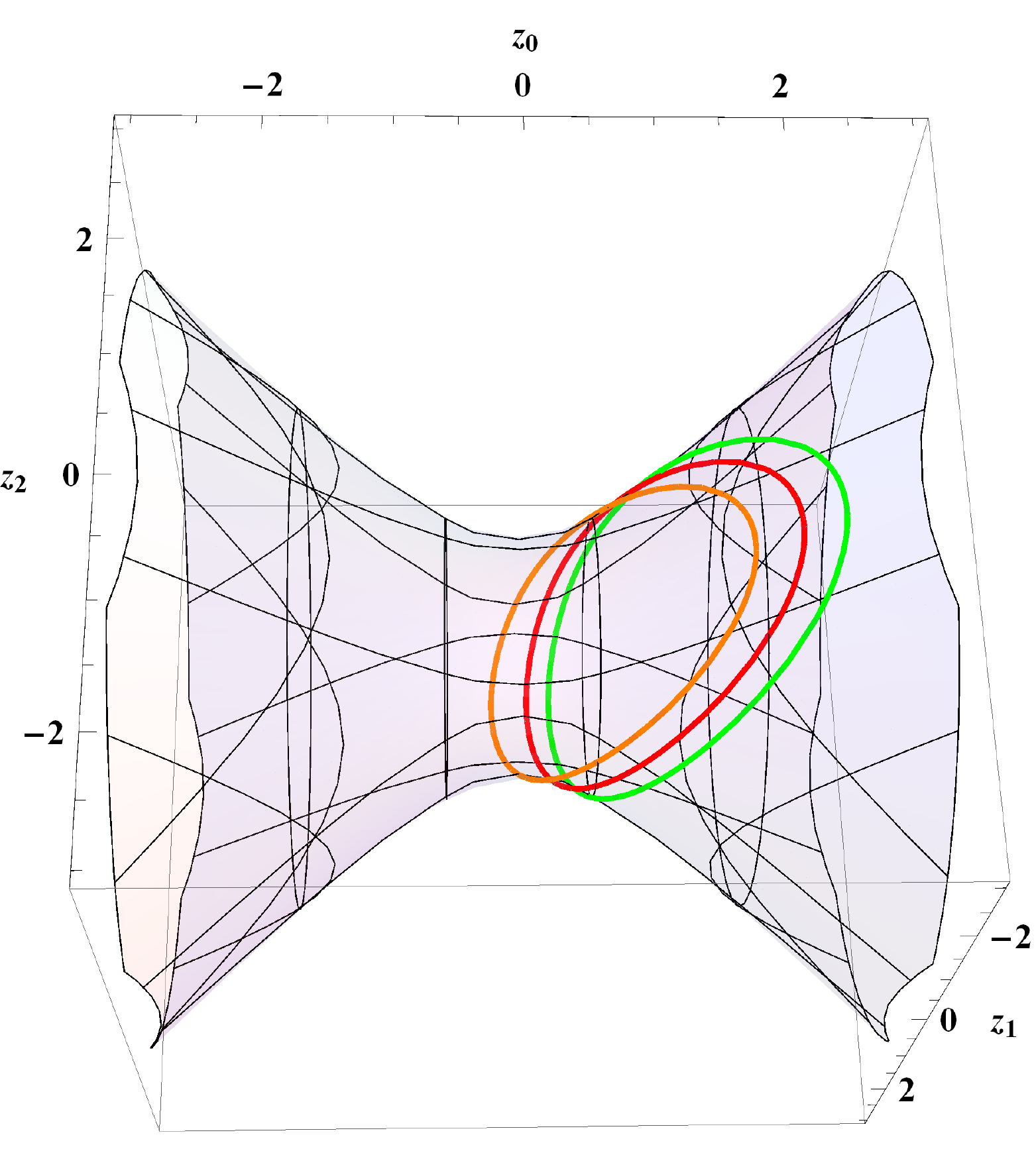}
\caption{The paths of motion for $\epsilon=0.8$ and $p=1.5\,,\,2\,,\,2.5$}
\end{figure}

Let us denote by $2a$ the length of the major axis of the ellipse and by $2c$ the focal distance. It is clear that
\begin{eqnarray}
\label{H-J1-33}
\tanh 2a &=& \tanh (\tau_{\max} + \tau_{\min}) = \frac{\tanh \tau_{\min} + \tanh \tau_{\max}}
{1+ \tanh \tau_{\min} \cdot \tanh \tau_{\max}}
= - \frac{\alpha}{R E},
\\[3mm]
\tanh 2c &=& \tanh (\tau_{\max} - \tau_{\min}) = \frac{\tanh \tau_{\min} - \tanh \tau_{\max}}
{1- \tanh \tau_{\min} \cdot \tanh \tau_{\max}}
= \frac{\varepsilon (R)}{p- \coth 2a}.
\nonumber
\end{eqnarray}
Thus we get the result true also in Euclidean space, that the length of the major axis
of the ellipse depends only on the energy. The bounded orbits are shown in the Fig. 2.

We can easily calculate the period of the elliptic motion.  Using  equations (\ref{H-J6}), (\ref{H-J30})
and (\ref{H-J33}) and taking into account the last formula, we obtain
\begin{eqnarray}
\label{H-J33-001}
T &=&  \frac{R^2}{L} \int_{\tanh\tau_{min}}^{\tanh\tau_{max}}\frac{d X}{(1-X^2) \sqrt{- X^2  +
{2\alpha R}/{L^2} X +  {2R^2E}/{L^2} + 1}}
\nonumber\\[2mm]
&=& \frac{\pi R}{\sqrt{8|E|}}\left[\frac{1}{\sqrt{1+\alpha/R|E|}} + \frac{1}{\sqrt{1- \alpha/R|E|}}
\right].
\end{eqnarray}
Thus, the period $T$ depends only on energy. Using now equation (\ref{H-J1-33}) we can rewrite
equation (\ref{H-J33-001}) in the form
\begin{equation}
\label{H-J33-100}
T^2  =  \frac{\pi^2 R^{3}}{\alpha} \sinh a \cosh^3 a,
\end{equation}
which shows that the square of period depends only on the major axis of the orbit
({\it the third Kepler's law}). In the case of the minimal energy (see equation (\ref{H-J9})) we have
$\varepsilon =0$ and again find that the orbits are circles (see Fig.3).

\begin{figure}[H]
\centering
\includegraphics[width=7cm,height=7cm]{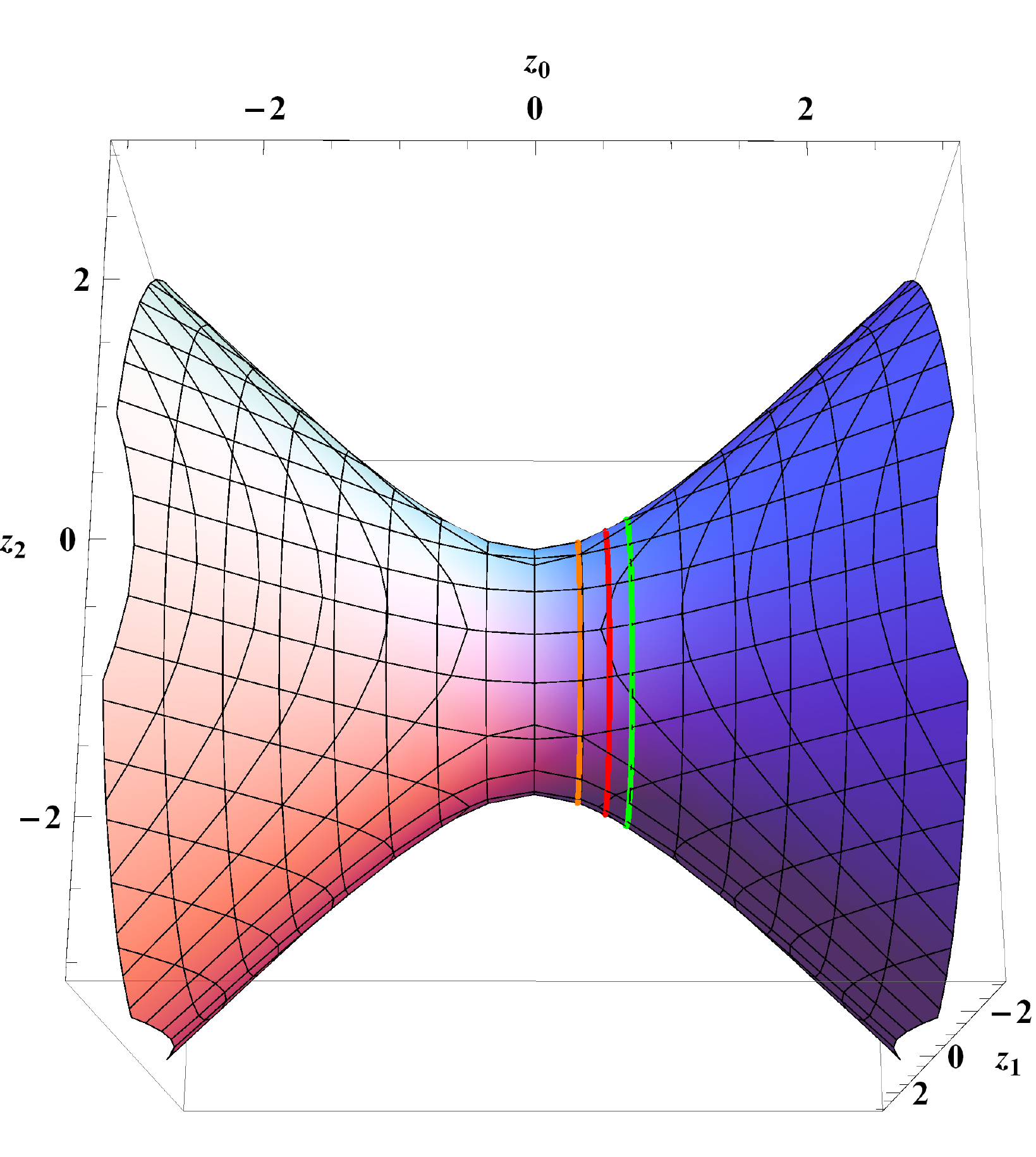}
\includegraphics[width=7cm,height=7cm]{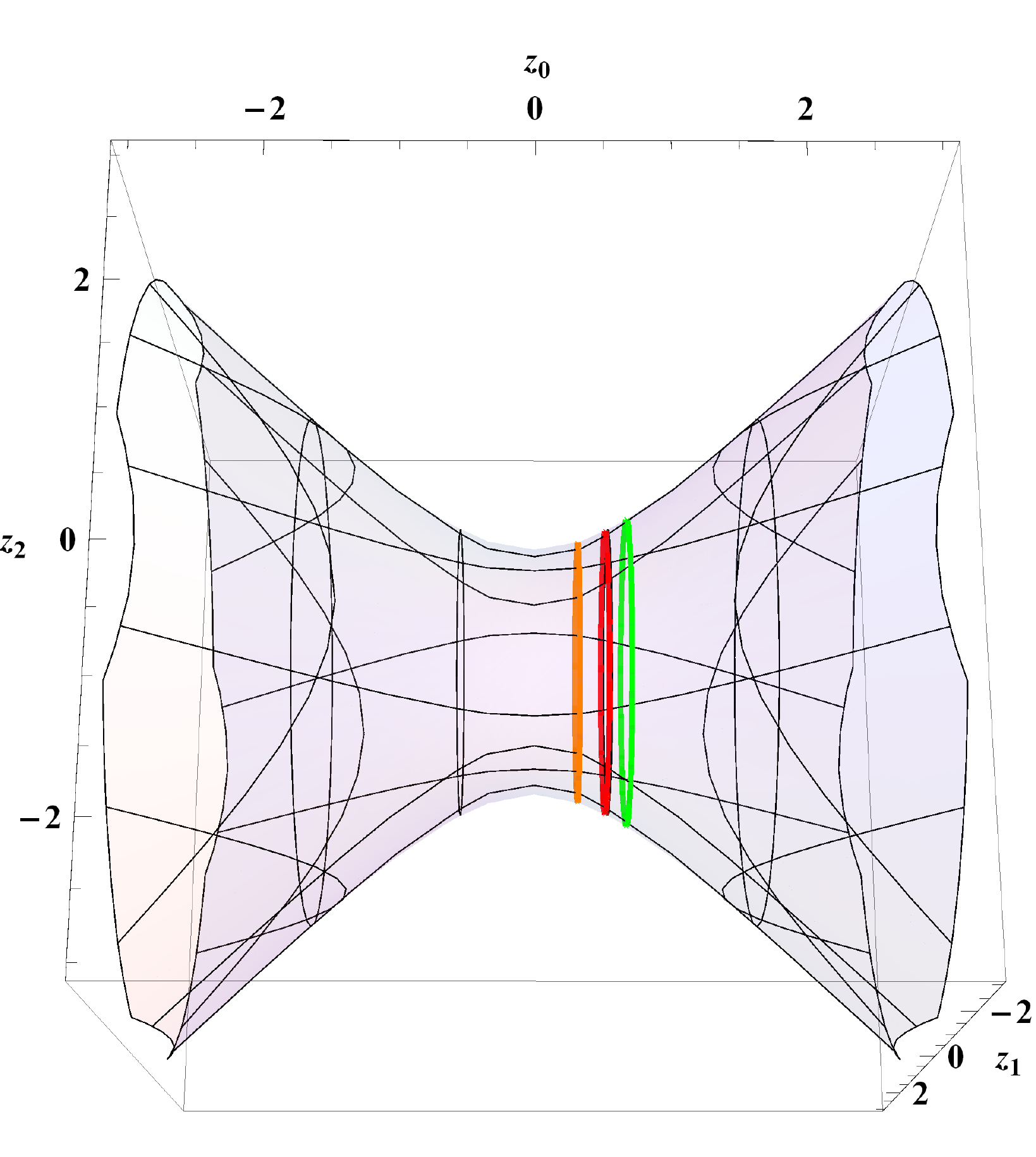}
\caption{The paths of otion for $\epsilon=0$ and $p=1.5\,,\,2\,,\,2.5$}
\label{c1-circle}
\end{figure}
\vspace{0.2cm}

{\bf B.}
For $E=-\alpha/R$ we have $\varepsilon (R) = 1$ and the particle moves from infinity $(\tau \sim \infty)$
at $\phi = \pi$, is reaches the turning point $\tau = \tanh^{-1}(2\alpha R/L^2-1)$ ( $\phi = 0$) and returns to
infinity. The path of the motion is a parabola (see Fig.4).
\begin{figure}[H]
\centering
\includegraphics[width=7cm,height=7cm]{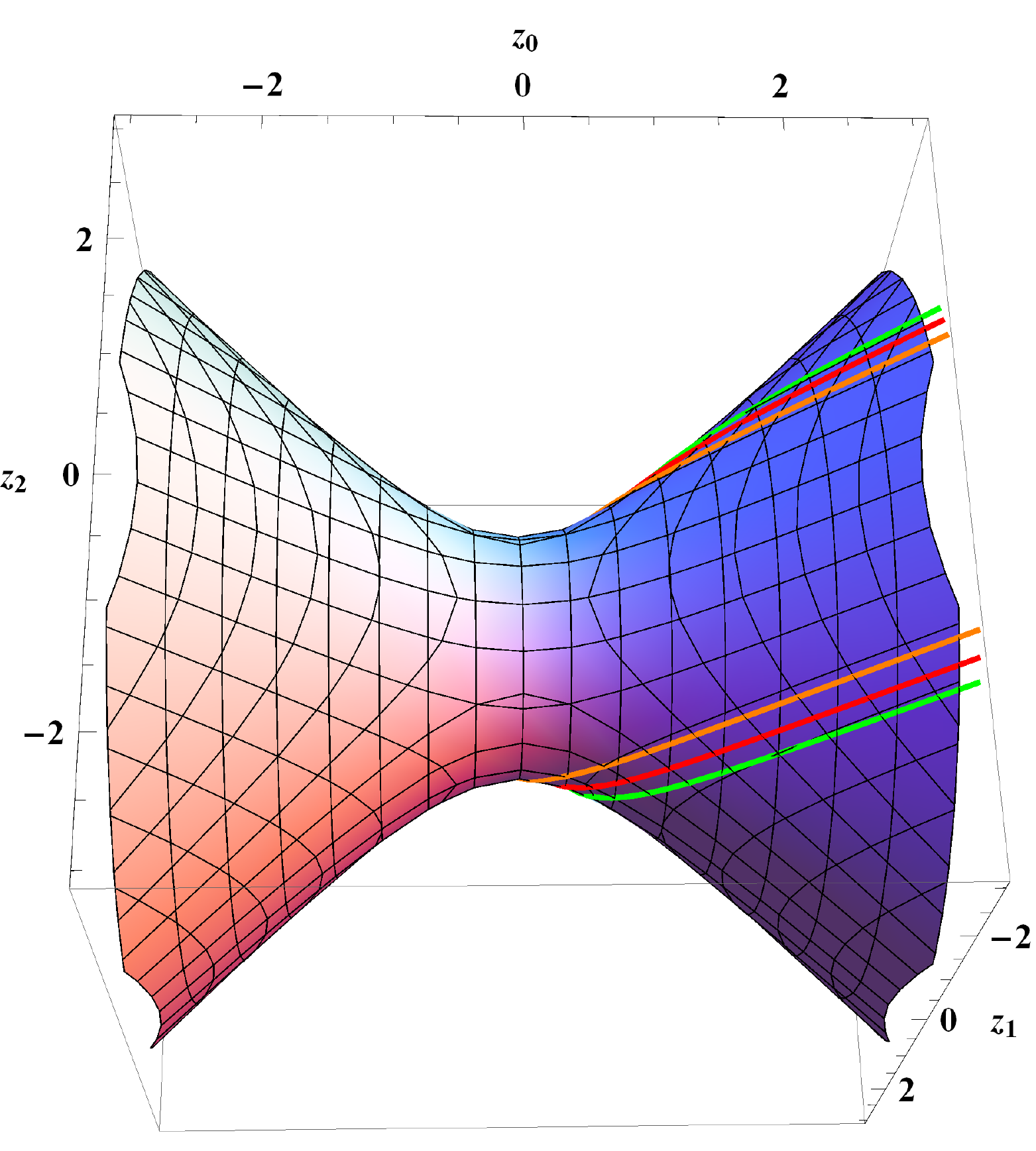}
\includegraphics[width=7cm,height=7cm]{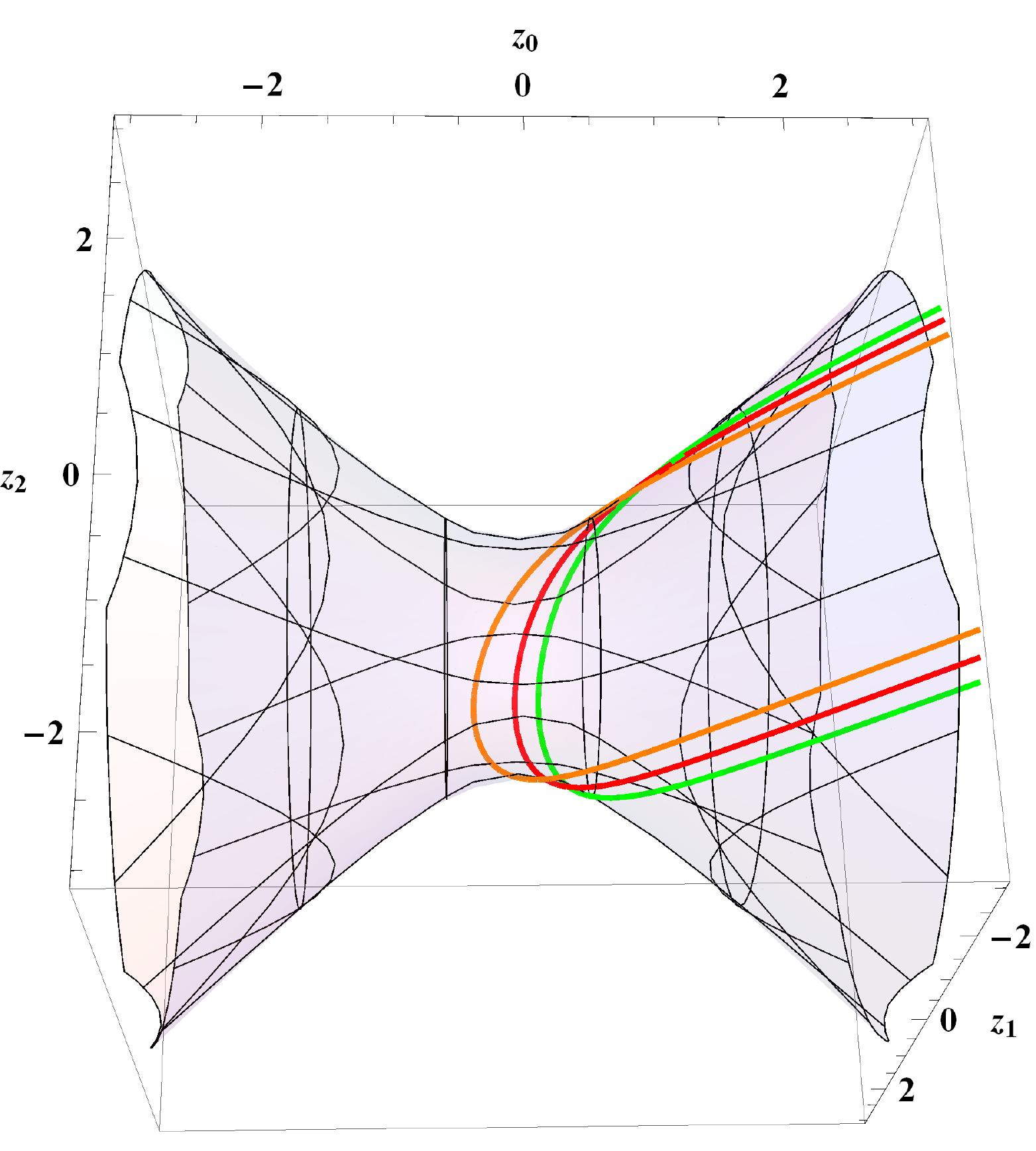}
\caption{The paths of motion for $\epsilon=1$ and $p=1.5\,,\,2\,,\,2.5$}
\end{figure}
\vspace{0.2cm}

{\bf C.} For $- \alpha/R < E < \alpha/R$, the path of the motion is a hyperbola. The particle
again moves from infinity at $\phi = \pi/2 - \sin^{-1}[\varepsilon (R)^{-1}]$, turns at the point
$\tau = \tanh^{-1}(\alpha R/L^2- \sqrt{(\alpha R/L^2)^2+2R^2E/L^2 +1})$  ($\phi = 0$) and goes back to
infinity (see Fig.5 and Fig. 6).
\begin{figure}[H]
\centering
\includegraphics[width=7cm,height=7cm]{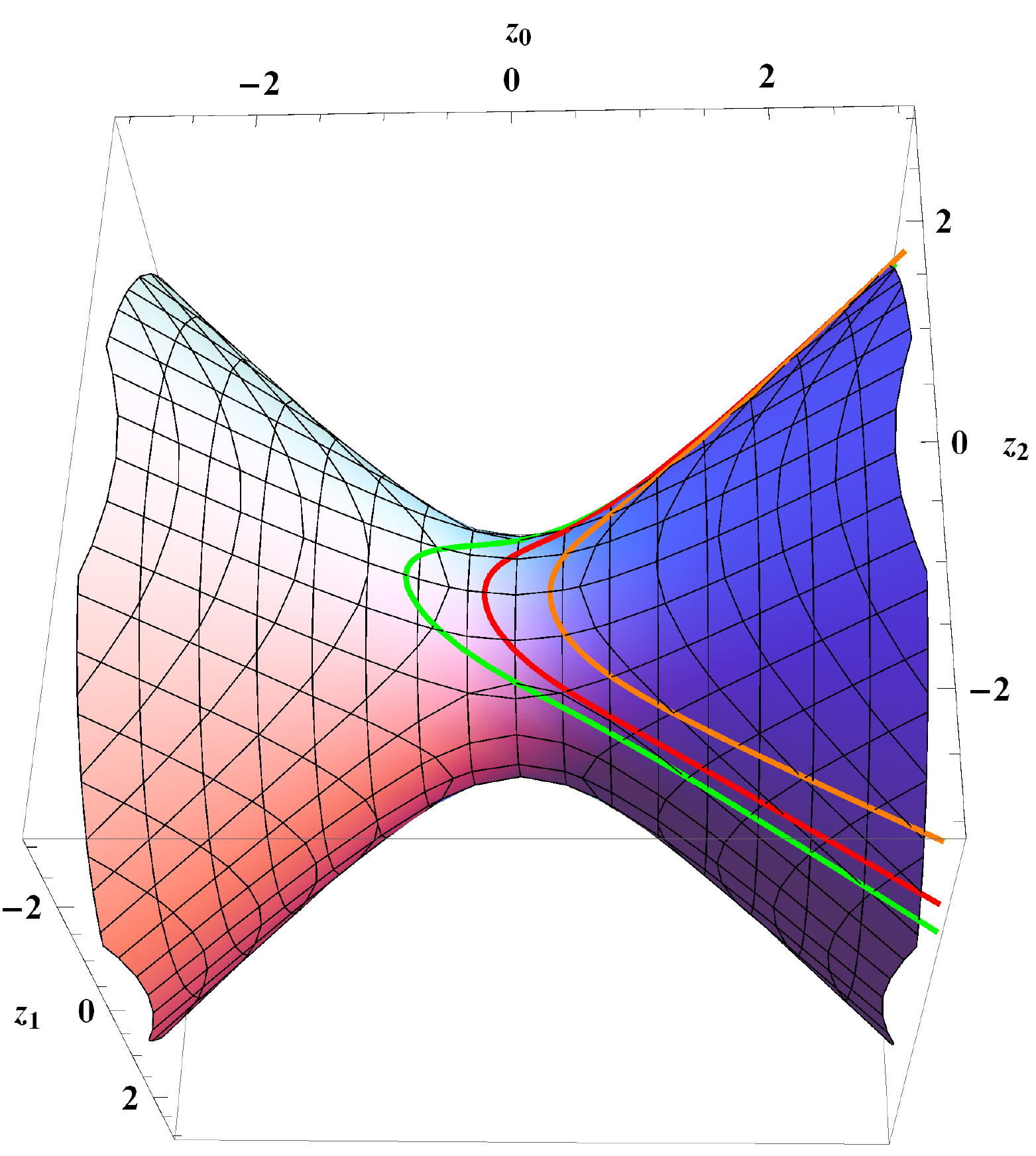}
\includegraphics[width=7cm,height=7cm]{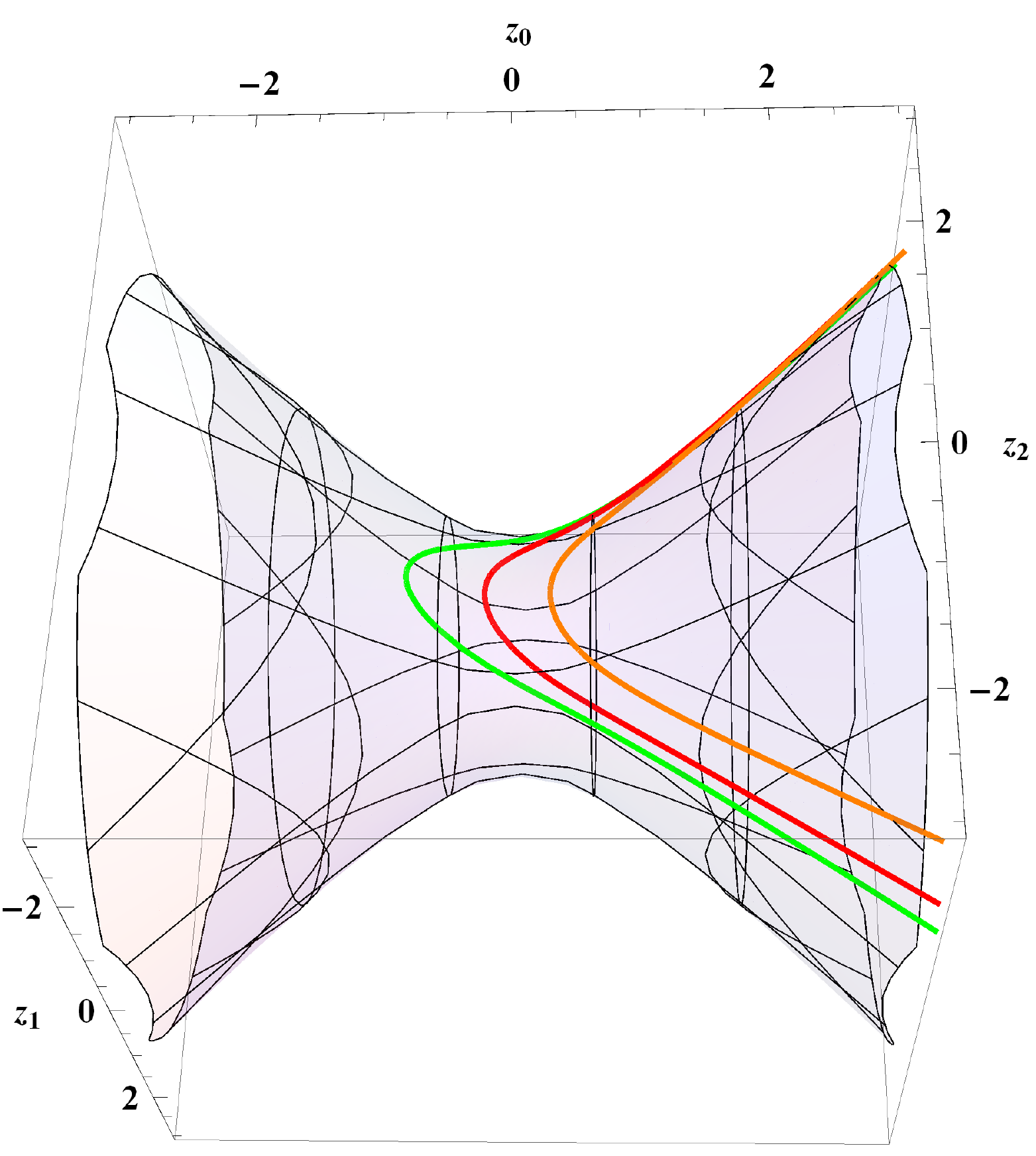}
\caption{The paths of motion for $L^2=1/2$ and $E=-1/2\,,\,0\,,\,1/2$ ($\alpha=R=1$)}
\end{figure}

\begin{figure}[H]
\centering
\includegraphics[width=7cm,height=7cm]{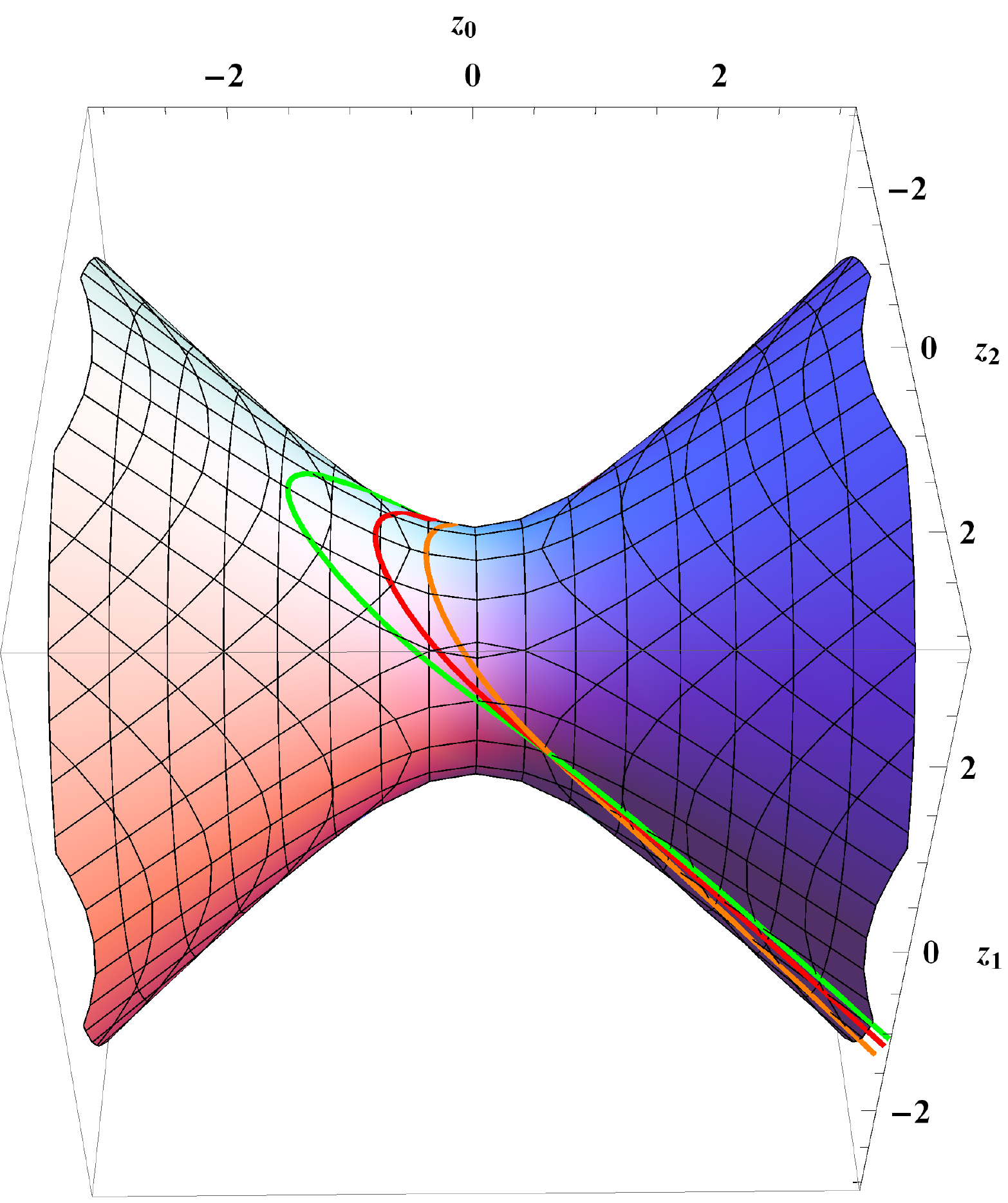}
\includegraphics[width=7cm,height=7cm]{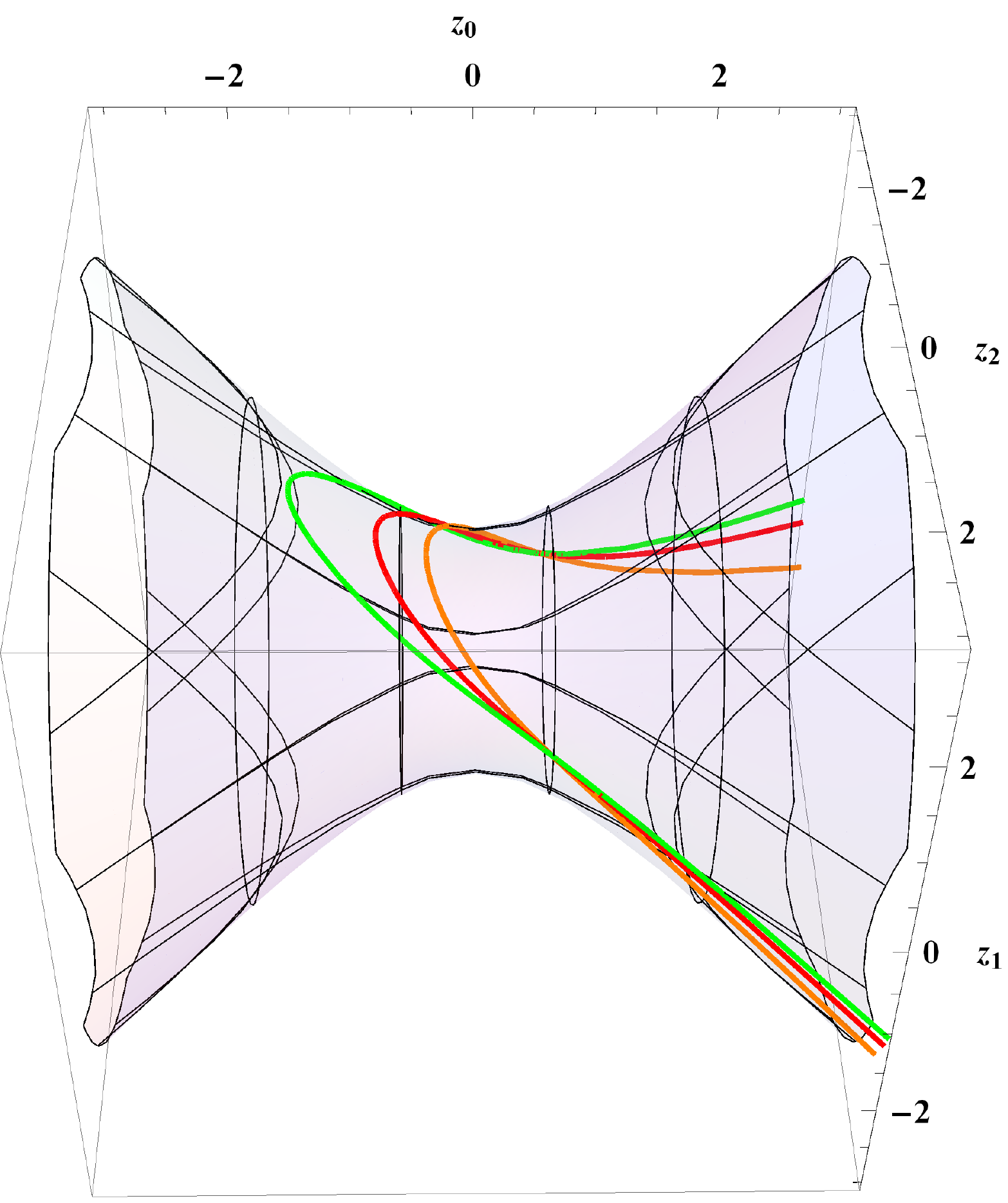}
\caption{The paths of motion for $L^2=2$ and $E=-1/2\,,\,0\,,\,1/2$ ($\alpha=R=1$)}
\end{figure}
\vspace{0.2cm}

{\bf D.}
For $E \geq \alpha/R$, $\tau \in (- \infty, \infty)$
(see Fig.7 and Fig.8).
\begin{figure}[H]
\centering
\includegraphics[width=7cm,height=7cm]{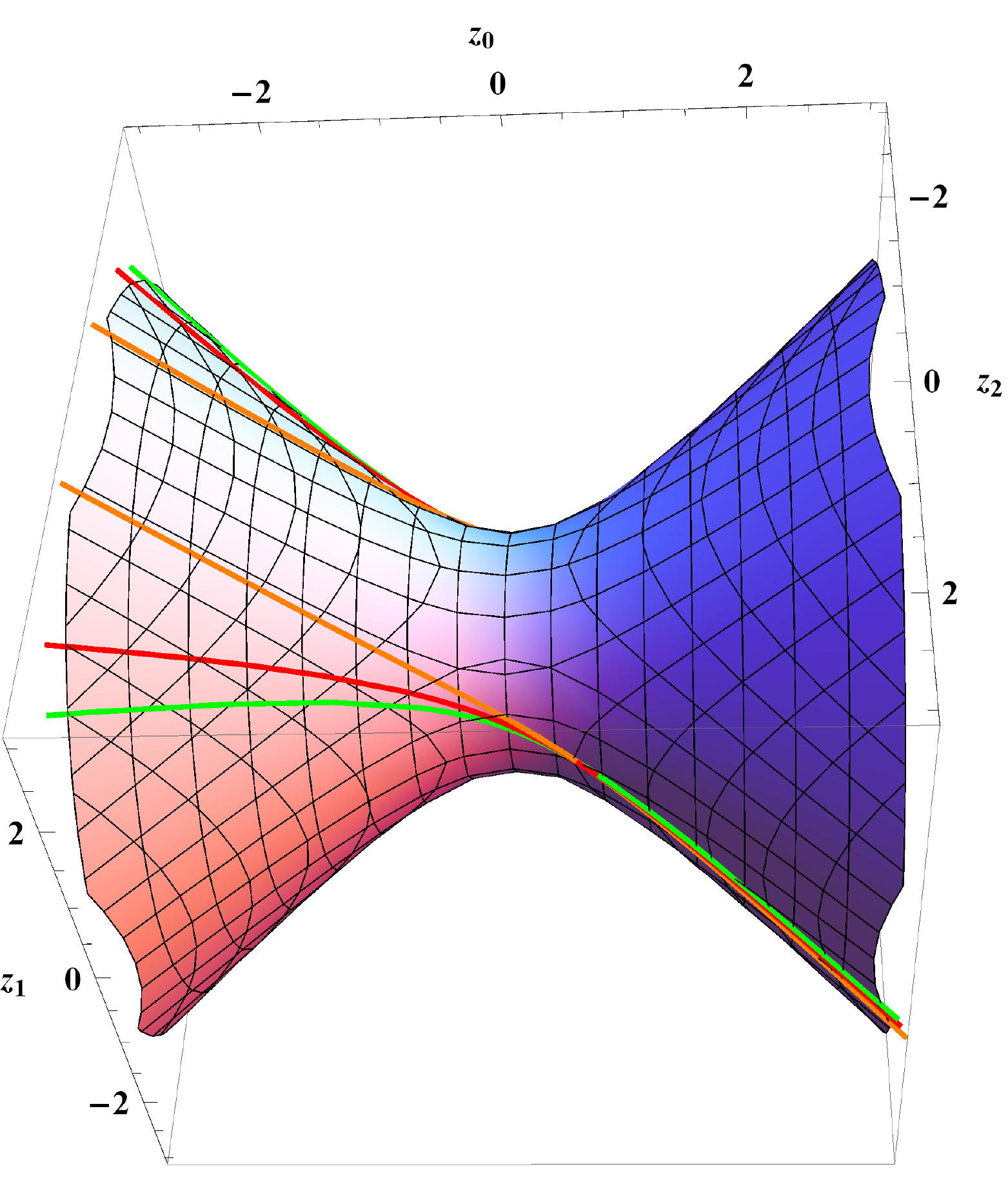}
\includegraphics[width=7cm,height=7cm]{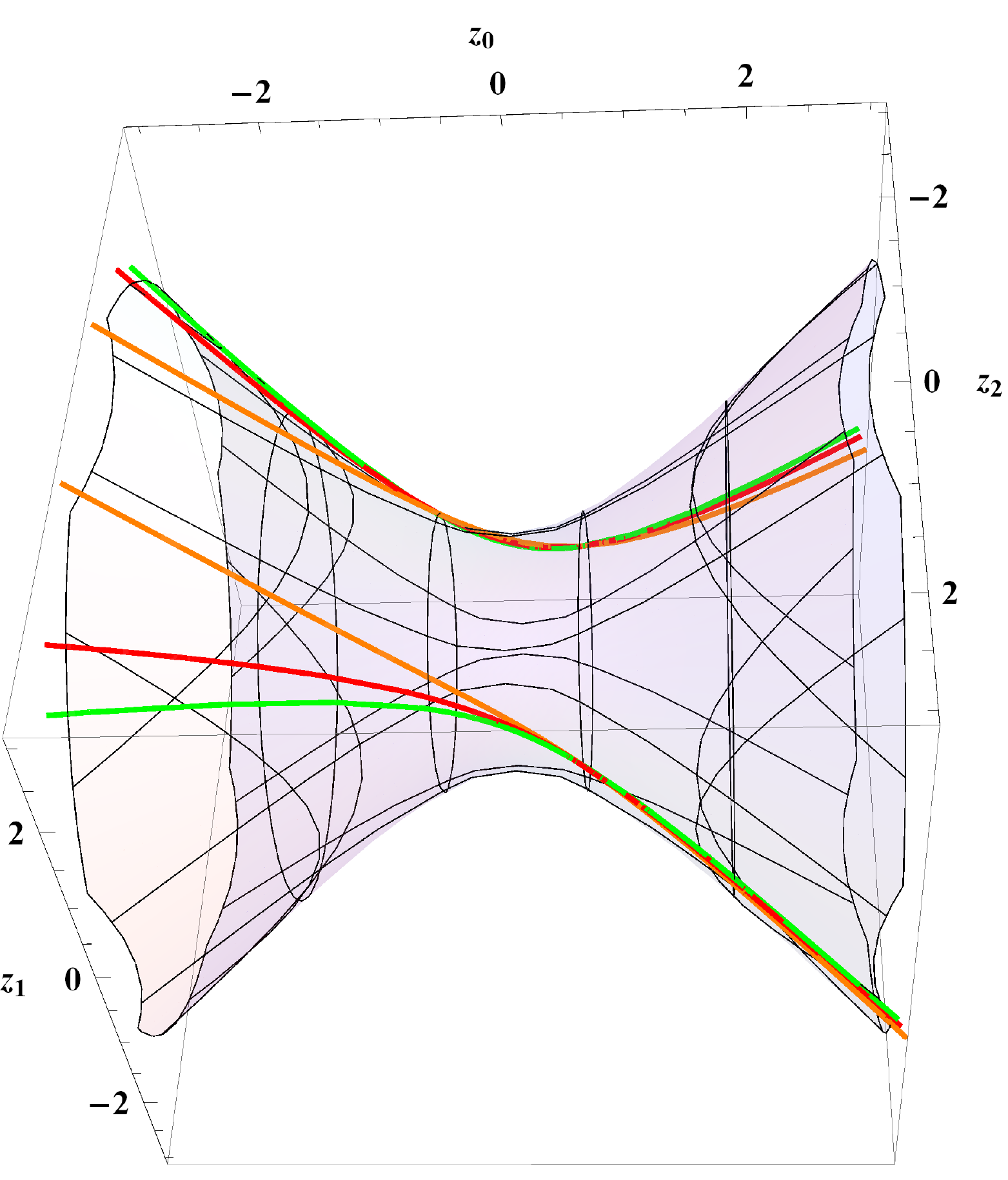}
\caption{The paths of motion for $L^2=2$ and $E=-1/2\,,\,0\,,\,1/2$ ($\alpha=R=1$)}
\end{figure}

\begin{figure}[H]
\centering
\includegraphics[width=7cm,height=7cm]{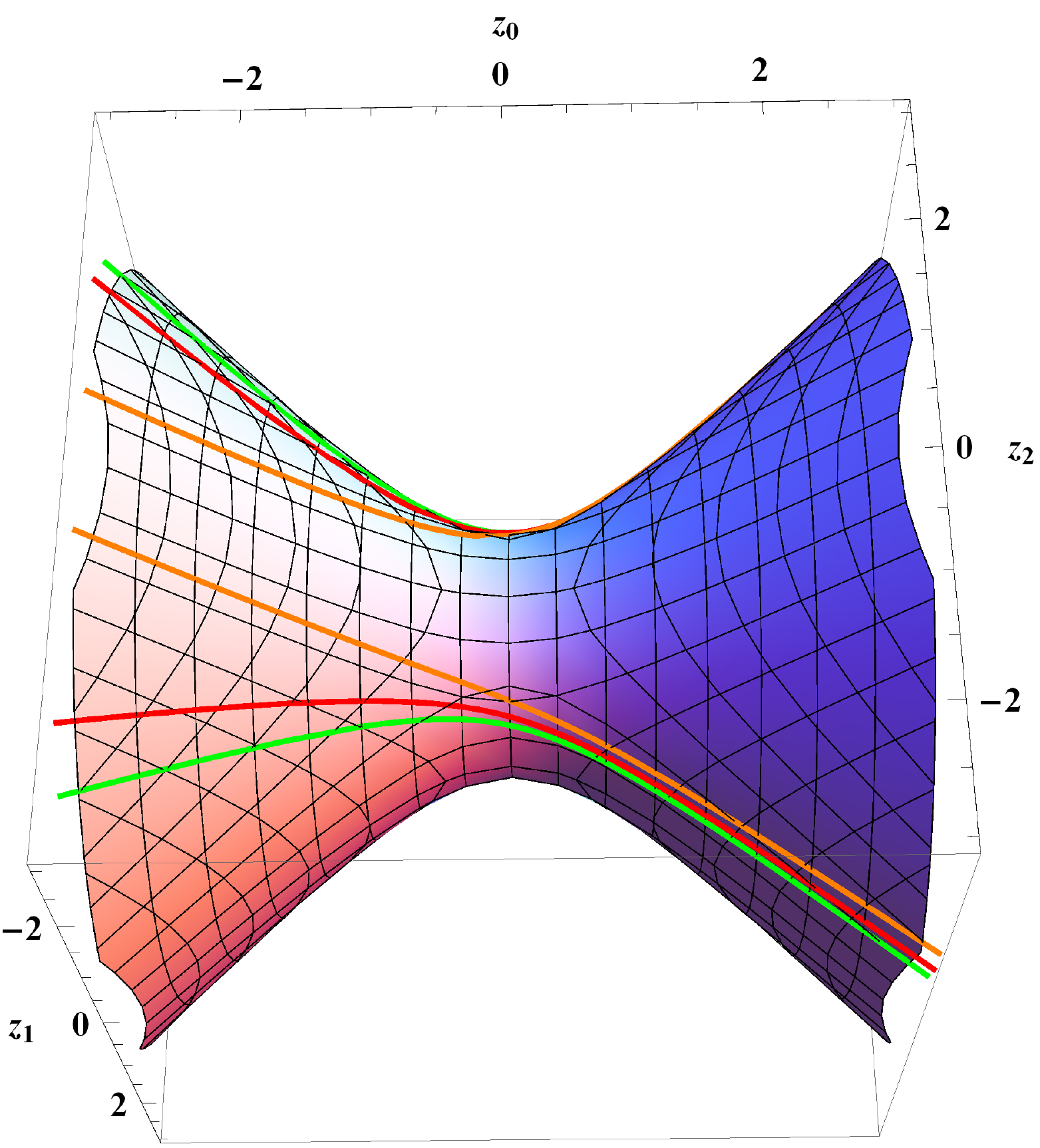}
\includegraphics[width=7cm,height=7cm]{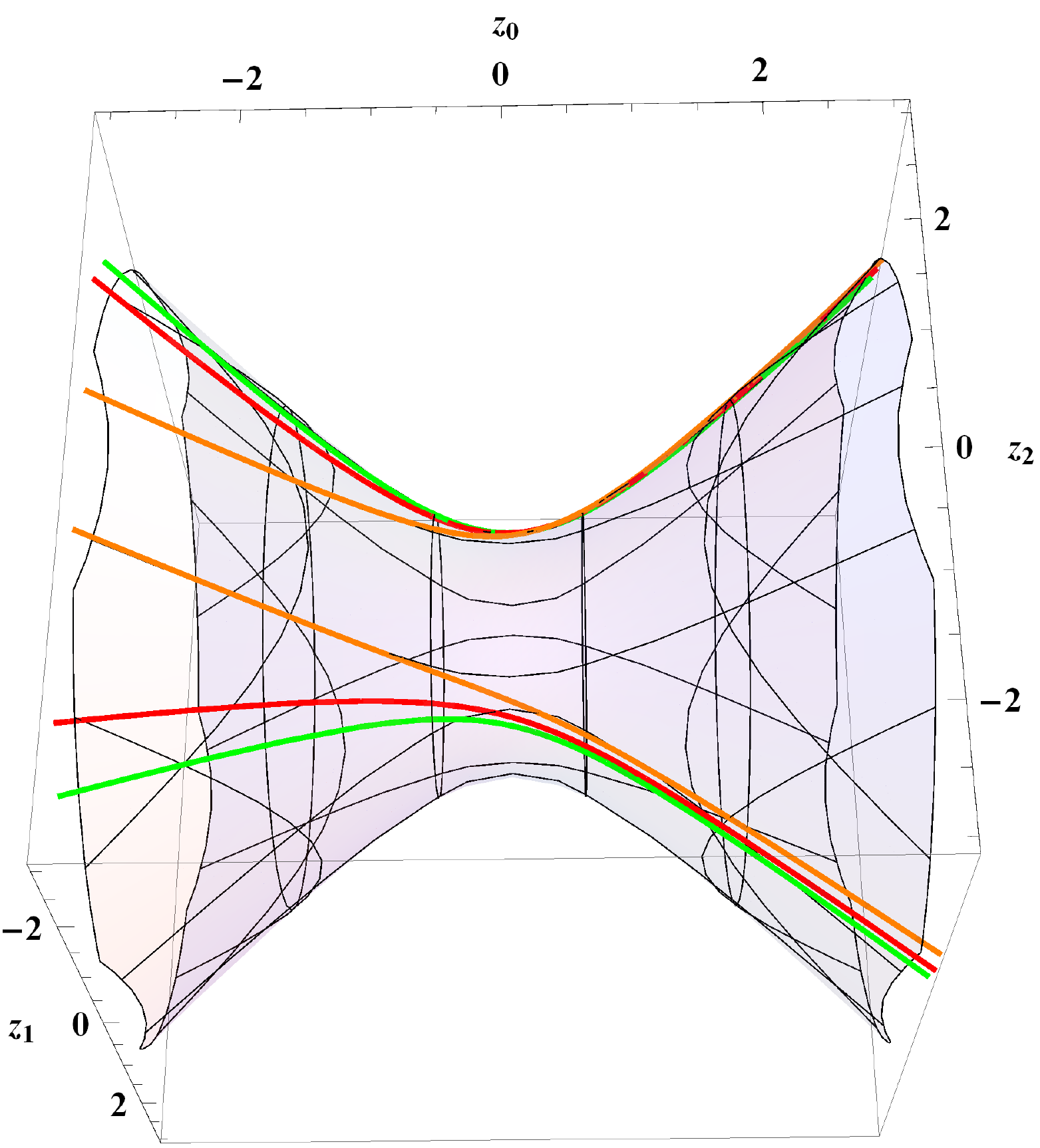}
\caption{The paths of motion for $L^2=1/2$ and $E=-1/2\,,\,0\,,\,1/2$ ($\alpha=R=1$)}
\end{figure}

\vspace{0.3cm}
\noindent
\newpage
{\bf\large Acknowledgments}

\vspace{0.2cm}
\noindent
The work of Yu.A.K., L.G.M, V.S.O., and G.S.P. was supported in part by the Armenian-Belarusian grants Nos. 13RB-035 and Ph14ARM-029 from SCS and FFR.

%-----------------------------------------------------------------------------

\end{document}